\documentstyle[12pt]{article}

\def\bfkappa{\mathop{\mbox{\boldmath $\kappa$}}}

\newcommand{\be}{\begin{equation}}
\newcommand{\ee}{\end{equation}}
\newcommand{\bi}[1]{\vspace{-3mm} \bibitem{#1}}

\begin{document}
%%%%%%%%%%%%%%%%%%%%%%%%%%%%%%%%%%%%%%%%%%%%%%%%%%%%%%%%%%%%%%%%%%%%%%
%\today

\begin{center}
{\Large \bf Dynamics with Low-Level Fractionality}
\vskip 5 mm

{\large \bf Vasily E. Tarasov$^{1,2}$ and George M. Zaslavsky$^{1,3}$ } \\

\vskip 3mm

{\it $1)$ Courant Institute of Mathematical Sciences, New York University \\
251 Mercer Street, New York, NY 10012, USA  }\\ 
{\it $2)$ Skobeltsyn Institute of Nuclear Physics, \\
Moscow State University, Moscow 119992, Russia } \\
{\it $3)$ Department of Physics, New York University, \\
2-4 Washington Place, New York, NY 10003, USA } \\
%%{E-mail: tarasov@theory.sinp.msu.ru}}
\end{center}

\vskip 11 mm

\begin{abstract}
The notion of fractional dynamics is related to equations 
of motion with one or a few terms with derivatives 
of a fractional order.
This type of equation appears in the description of chaotic
dynamics, wave propagation in fractal media, and field theory.
For the fractional linear oscillator the physical meaning
of the derivative of order $\alpha<2$ is dissipation.
In systems with many spacially coupled elements (oscillators) 
the fractional derivative, along the space coordinate, 
corresponds to a long range interaction.
We discuss a method of constructing a solution using 
an expansion in $\varepsilon=n-\alpha$ with small $\varepsilon$
and positive integer $n$. 
The method is applied to
the fractional linear and nonlinear oscillators
and to fractional Ginzburg-Landau or parabolic equations.
\end{abstract}

\vskip 3 mm
{\small 


%%%45.10.Hj 	Perturbation and fractional calculus methods
%%%45.05.+x 	General theory of classical mechanics of discrete systems
%%%45.50.-j 	Dynamics and kinematics of a particle and a system of particles
%%%05.45.Df 	Fractals
%%%05.40.Fb 	Random walks and Levy flight
%%%05.40.-a 	Fluctuation phenomena, random processes, noise, and Brownian motion
%%%05.60.-k 	Transport processes

\vskip 3 mm


\vskip 11 mm

\section{Introduction}

It became clear in the last decade that many physical processes 
can be adequately described by equations that consist of 
derivatives of fractional order.
In a fairly short period of time, the list of such 
applications is long and the areas of applications are broad.
Even in a concise form, the applications include material
sciences \cite{Hilfer,C2}, chaotic dynamics \cite{Zaslavsky1},
quantum theory \cite{Laskin,Naber,Krisch,Goldfain}, 
physical kinetics \cite{Zaslavsky1,ZE,SZ,Zaslavsky7},
fluids and plasma physics \cite{CLZ,Nig1,Nig3,Plasma2005}, 
and many others physical topics
related to anomalous diffusion, wave propagation \cite{ZL}, and
transport theory (see more in reviews \cite{Zaslavsky1,MK}).
Since the fractional calculus has a fairly long history,
the approaches for solutions of corresponding equations
are also enormously rich.
Let us mention some of these approaches 
that are related more specifically to this paper:
probabilistic basis and interpretation of the 
fractional kinetics \cite{MonS,UZ2,SZ,WBG},
dissipative interpretation of the fractional derivative 
\cite{GM2,GM3,M,ZSE},
Green's function method \cite{MR,Podlubny}, etc. 

This paper is motivated by a lack of methods that permit
an explicit analysis of the equations with fractional derivatives
that describe some important physical processes.
Particularly, the difficulties are due to the absence
of characteristic scales (sometimes such processes are 
called multiscaling) and the difficulty of direct estimates.
In typical physical situations the "level of fractality" is low, 
i.e. the order of fractional derivatives $\alpha$ of 
the corresponding terms of the dynamical equations
slightly deviates from an integer value $n$
(in the considered cases $n=1$ or 2).
This recalls a possibility to use an expansion
over the small parameter $\varepsilon=n-\alpha$
that we call an $\varepsilon$-expansion.
We develop a construction that is applied to 
linear and nonlinear fractional oscillators.
Particularly, for the linear fractional oscillator (LFO)
the obtained expansion can be compared to the exact one.
As a more complicated example, we consider a solution of 
the fractional Ginzburg-Landau equation (FGL)
introduced in \cite{Zaslavsky6,Physica2005} (see also \cite{Mil}).

The basic description of the $\varepsilon$-expansion is given in Sec. 2
for different fractional derivatives.
In Sec. 3, we consider examples of the $\varepsilon$-expansion 
application to linear and nonlinear oscillators.
In Sec. 4, we consider large $t$ asymptotics ($t\rightarrow \infty$).
Application of the $\varepsilon$-expansion to the Landau-Ginzburg
or to the parabolic nonlinear equation is given in Sec. 5.
Some technical details are the Appendices 1, 2.

%%%%%%%%%%%%%%%%%%%%%%%%%%%%%%%%%%%%%%%%%%%%%%%%%%%%%%%%%
\section{Description of the $\varepsilon$-expansion}

\subsection{Caputo fractional derivative of order $\alpha=2-\varepsilon$}

The fractional derivative has different definitions \cite{SKM,OS}, 
and exploiting any of them depends on the kind of 
the problems, initial (boundary) conditions, and 
the specifics of the considered physical processes.
The classical definition is the so-called Riemann-Liouville
derivative \cite{SKM,OS,GS}
\[
_a{\cal D}^{\alpha}_{t}f(x)=
\frac{1}{\Gamma(n-\alpha)} \frac{\partial^n}{\partial x^n}
\int^{x}_{a} \frac{f(z) dz}{(x-z)^{\alpha-n+1}},
\]
\be \label{D2}
_t{\cal D}^{\alpha}_{b}f(x)=
\frac{(-1)^n}{\Gamma(n-\alpha)} \frac{\partial^n}{\partial x^n}
\int^{b}_x \frac{f(z) dz}{(z-x)^{\alpha-n+1}} ,
\ee
where $n-1<\alpha<n$.
Due to reasons, concerning the initial conditions,
it is more convenient to use the Caputo fractional derivatives
\cite{C2,C1,C3}.
Its main advantage is that the initial conditions take the same
form as for integer-order differential equations.

The left Caputo fractional derivative \cite{C2,C1,C3} is defined by the equation
\be \label{1}
D^{\alpha}f(t)=\ _0^CD^{\alpha}_tf(t)=
\frac{1}{\Gamma(n-\alpha)} 
\int^t_0 \frac{f^{(n)}(\tau)}{(t-\tau)^{\alpha+1-n}} d \tau ,
\ee
where $n-1 < \alpha < n$, and $f^{(n)}(\tau)=d^n f(\tau)/d\tau^n$. 
For $n=2$, 
\be \label{2}
D^{2-\varepsilon}f(t)=\frac{1}{\Gamma(\varepsilon)} 
\int^t_0 \frac{f^{(2)}(\tau)}{(t-\tau)^{1-\varepsilon}} d\tau ,
\quad 0<\varepsilon <1 .
\ee
This presentation is not convenient in the limit 
$\varepsilon \rightarrow 0$, since
\be \label{3}
\frac{1}{\Gamma(\varepsilon)}=\varepsilon+O(\varepsilon^2) ,
\ee
and we present (\ref{1}) and (\ref{2}) in the form \cite{Podlubny}:
\be \label{4}
D^{\alpha}f(t)=\frac{f^{(n)}(0) t^{n-\alpha}}{\Gamma(n-\alpha+1)} 
+\frac{1}{\Gamma(n-\alpha+1)} \int^t_0 f^{(n+1)}(\tau) (t-\tau)^{n-\alpha} d\tau ,
\quad n-1<\alpha \le n ,
\ee
or for $n=2$,  
\be \label{5}
D^{2-\varepsilon }f(t)=\frac{f^{(2)}(0) t^{\varepsilon}}{\Gamma(1+\varepsilon)}  
+\frac{1}{\Gamma(1+\varepsilon)} \int^t_0 f^{(3)}(\tau) (t-\tau)^{\varepsilon} 
d \tau.
\ee
Let us consider first the case 
\be \label{20'}
\varepsilon \ t \ll 1. 
\ee
We can use the expansion
\[ \frac{1}{\Gamma(1+\varepsilon)} (t-\tau)^{\varepsilon}=
\frac{1}{\Gamma(1+\varepsilon)} e^{\varepsilon \ln(t-\tau)}= \] 
\be
\label{6}
=1+\varepsilon \left( \gamma+\ln(t-\tau) \right)+ \varepsilon^2\left( 
\frac{1}{2} \ln^2(t-\tau)+
\gamma \ln(t-\tau)+\frac{\gamma^2}{2}-\frac{\pi^2}{12}
\right)+...,
\ee
where $\tau<t$, and $\gamma=0.5772156649...$ is a constant.
As the result, we obtain
\be \label{7}
D^{2-\varepsilon }f(t)=f^{(2)}(t)  +\varepsilon \left( f^{(2)}(0) \ln(t)+ 
\gamma f^{(2)}(t)+ \int^t_0 f^{(3)}(\tau) \ln(t-\tau ) d \tau \right) + ...
\ee 
that expresses the fractional derivative in a form of
the perturbation to the second derivative, when $\varepsilon\ll1$:
\be \label{8}
D^{2-\varepsilon }f(t)=f^{(2)}(t)  +\varepsilon D^2_1 f(t) + ... . 
\ee
where $D^2_1$ is defined as:
\be \label{res1}
D^2_1 f(t)=f^{(2)}(0) \ln(t)+ 
\gamma f^{(2)}(t)+ \int^t_0 f^{(3)}(\tau) \ln(t-\tau ) d \tau .
\ee
Note that, the limit $\varepsilon \rightarrow 0$ in equation (\ref{7}) gives
the correct expansion
\[ \lim_{\varepsilon \rightarrow 0} D^{2-\varepsilon }f(t)=f^{(2)}(t) . \]
Another useful comments is that the 3d derivative $f^{(3)}(t)$ should
exist in order to use the correction of order $\varepsilon$ in (\ref{7}).

%%%\subsection{Simple example}

As a simple example of application of formula (\ref{7}), consider 
the $\alpha$-derivative of $t^2$ and $t^3$:
\[
D^{2-\varepsilon } t^2=2  +2 \varepsilon \gamma + ... 
\]
\be \label{E2} 
D^{2-\varepsilon } t^3=6t  +\varepsilon 
[6 ( \gamma-1) t+6 t \ln(t)] + ... 
\ee
From the well-known relation
\be \label{E3}
D^{\alpha} t^{\beta} =\frac{\Gamma(\beta+1)}{\Gamma(\beta-\alpha+1)} t^{\beta-\alpha},
\ee 
we get the exact results
\[
D^{2-\varepsilon} t^2 =
\frac{2}{\Gamma(1+\varepsilon)} t^{\varepsilon},
\]
\be \label{E5}
D^{2-\varepsilon} t^3 =
\frac{6}{\Gamma(2+\varepsilon)} t^{1+\varepsilon} .
\ee
Expansions of (\ref{E5})
for $\varepsilon \ t\ll1$ coincide with (\ref{E2}).

\subsection{Caputo fractional derivative of order  $\alpha=1-\varepsilon$.}

To consider the limit $\alpha \rightarrow 1$, let us put $n=1$ in (\ref{4}):
\be \label{ss10}
D^{1-\varepsilon }f(t)=\frac{f^{(1)}(0) t^{\varepsilon}}{\Gamma(1+\varepsilon)} 
 +\frac{1}{\Gamma(1+\varepsilon)} \int^t_0 f^{(2)}(\tau) 
(t-\tau)^{\varepsilon} d \tau.
\ee
Similarly to $n=2$, consider the case $\varepsilon \ t \ll1$, 
and use the expansion (\ref{6}). It gives 
\be \label{ss11}
D^{1-\varepsilon }f(t)=f^{(1)}(t)  +\varepsilon D^1_1 f(t) + ... ,
\ee
where $D^1_1 f(t)$ is 
\be \label{ss12}
D^1_1 f(t)= f^{(1)}(0) \ln(t)+ \gamma f^{(1)}(t)+ 
\int^t_0 f^{(2)}(\tau) \ln(t-\tau ) d \tau .
\ee
Note that, if we consider the limit $\varepsilon \rightarrow 0$, we get
\be \label{ss1-3}
\lim_{\varepsilon \rightarrow 0} D^{1-\varepsilon }f(t)=f^{(1)}(t) . 
\ee

%%%%%%%%%%%%%%%%%%%%%%%%%%%%%%%%%%%%%%%%%%%%%%%%%%%%%
\subsection{Riesz fractional derivatives of order $\alpha=2-\varepsilon$}

The Riesz fractional derivative of order $\alpha$ is defined by 
\be \label{R1}
\frac{d^{\alpha}}{d |x|^{\alpha}} f(x)=
-\frac{1}{2 \cos(\pi \alpha /2)} 
\left({\cal D}^{\alpha}_{+}f(x) +{\cal D}^{\alpha}_{-}f(x)\right) ,
\ee
where $\alpha\not=1,3,5...$, and
${\cal D}^{\alpha}_{\pm}$ are Riemann-Liouville fractional derivatives
\[
{\cal D}^{\alpha}_{+}f(x)=
\frac{1}{\Gamma(n-\alpha)} \frac{\partial^n}{\partial x^n}
\int^{x}_{-\infty} \frac{f(z) dz}{(x-z)^{\alpha-n+1}},
\]
\be \label{R3}
{\cal D}^{\alpha}_{-}f(x)=
\frac{(-1)^n}{\Gamma(n-\alpha)} \frac{\partial^n}{\partial x^n}
\int^{\infty}_x \frac{f(z) dz}{(z-x)^{\alpha-n+1}} .
\ee
Substitution of Eqs. (\ref{R3}) into Eq. (\ref{R1}) gives
\be \label{R4}
\frac{d^{\alpha}}{d |x|^{\alpha}} f(x)=
-\frac{1}{2 \cos(\pi \alpha /2) \Gamma(n-\alpha)} 
\frac{\partial^n}{\partial x^n} 
\left(
\int^{x}_{-\infty} \frac{f(z) dz}{(x-z)^{\alpha-n+1}}+
\int^{\infty}_x \frac{f(z) dz}{(z-x)^{\alpha-n+1}}
\right) .
\ee
For the order $\alpha=2-\varepsilon$, 
\be \label{R5}
\frac{d^{\alpha}}{d |x|^{\alpha}} f(x)=
\frac{1}{2 \cos(\pi \varepsilon /2) \Gamma(\varepsilon)} 
\frac{\partial^2}{\partial x^2} 
\left(
\int^{x}_{-\infty} \frac{f(z) dz}{(x-z)^{1-\varepsilon}}+(-1)^n
\int^{\infty}_x \frac{f(z) dz}{(z-x)^{1-\varepsilon}}
\right) .
\ee
Using the condition $f(\pm \infty)=0$, we get 
\[
\int^{x}_{-\infty} \frac{f(z) dz}{(x-z)^{1-\varepsilon}}=
\int^{x}_{-\infty} f(z) (x-z)^{\varepsilon-1} dz= 
\frac{1}{\varepsilon} \int^{x}_{-\infty} 
f^{(1)}(z) (x-z)^{\varepsilon} dz ,
\]
\be \label{R7}
\int^{\infty}_x \frac{f(z) dz}{(z-x)^{1-\varepsilon}}=
\int^{\infty}_x f(z) (z-x)^{\varepsilon-1} dz=
-\frac{1}{\varepsilon} \int^{\infty}_{x} 
f^{(1)}(z) (z-x)^{\varepsilon} dz .
\ee
These relations and equation 
$\varepsilon \Gamma(\varepsilon)=\Gamma(1+\varepsilon)$ 
allows us to present (\ref{R4}) in the form
\be \label{R8}
\frac{d^{\alpha}}{d |x|^{\alpha}} f(x)=
\frac{1}{2 \cos(\pi \varepsilon /2) \Gamma(1+\varepsilon)} 
\frac{\partial^2}{\partial x^2} \left(
\int^{x}_{-\infty} f^{(1)}(z) (x-z)^{\varepsilon} dz-
\int^{\infty}_x f^{(1)}(z) (z-x)^{\varepsilon} dz
\right) .
\ee
For the case $\varepsilon \ x \ll1$, we use
\be \label{R9}
\frac{(x-z)^{\varepsilon}}{cos(\pi \varepsilon /2) \Gamma(1+\varepsilon)}
=1+\varepsilon[\gamma +\ln(x-z)]+... 
\ee
to obtain
\be \label{R11}
\frac{d^{2-\varepsilon}}{d |x|^{2-\varepsilon}} f(x)=
%%%\frac{d^2 f(x)}{dx^2}
f^{(2)}(x)+\varepsilon D^2_{1 R} f(x)+... ,
\ee
where
\be \label{R12}
D^2_{1 R} f(x)=\gamma f^{(2)}(x) +
\frac{1}{2} \int^{\infty}_{0} 
\left[ f^{(3)}(x-y)-f^{(3)}(x+y)\right] \ln(y)dy .
\ee

A general comments to this section is that all results (\ref{res1}), 
(\ref{ss12}) and (\ref{R12}) are valid under restriction $\varepsilon t\ll 1$.
The expansion of Riemann-Liouville derivative
is considered in Appendix 1. 
For the asymptotic $t \rightarrow \infty$, and 
\be \label{33'}
\varepsilon t \gg 1 
\ee
another expansion is necessary.
The corresponding results will be considered in section 4.

%%%%%%%%%%%%%%%%%%%%%%%%%%%%%%%%%%%%%%%
\section{Linear and nonlinear fractional oscillator}

\subsection{$\varepsilon$-expansion for linear oscillator}

A linear fractional oscillator (LFO) 
is defined by the equation
\be  \label{11}
D^{\alpha} x(t)+\omega^2 x(t)=0 ,
\ee
where $\omega$ is dimensionless "frequency", 
$\alpha=2-\varepsilon$,  $0<\varepsilon \ll1$, and 
$D^{2-\varepsilon}$ is Caputo fractional derivative
that allows us to use the usual initial conditions \cite{Podlubny}.
LFO is an object of numerous investigations 
\cite{GM2,GM3,M,ZSE,Stanislavsky,AHC,AHEC,T,RR}
because of different applications.

The exact solution of Eq. (\ref{11}) for $1<\alpha<2$ is \cite{GM2,GM3}:
\be \label{11a}
x(t)=x(0)E_{\alpha,1}(-\omega^2 t^{\alpha})+
t x^{\prime}(0) E_{\alpha,2}(-\omega^2 t^{\alpha}),
\ee
where 
\be \label{11b}
E_{\alpha,\beta}(z)=\sum^{\infty}_{k=0} 
\frac{z^k}{\Gamma(\alpha k+\beta)} 
\ee
is the generalized two-parameter Mittag-Leffler function, and 
\be \label{11c}
E_{\alpha}(z)=E_{\alpha,1}(z)=\sum^{\infty}_{k=0} 
\frac{z^k}{\Gamma(\alpha k+1)} 
\ee
is one-parameter, or simply, Mittag-Leffler function \cite{ML1,ML2}. 

Existence of an exact solution of (\ref{11}) 
in the form (\ref{11a}) permits a comparison 
with the $\varepsilon$-expansion
\be \label{12}
x(t)=x_0(t)+\varepsilon x_1(t)+... 
\ee
constructed below.
The equation for $x_0(t)$ and the corresponding initial conditions are
\be  \label{1-3}
x^{\prime \prime}_0(t)+\omega^2 x_0(t)=0, 
\quad x_0(0)=x(0),  \quad x^{\prime}_0(0)=x^{\prime}(0),
\ee
with the solution 
\be \label{14}
x_0(t)=x(0) \cos(\omega t)+(x^{\prime}(0)/\omega) \sin(\omega t) .
\ee
The equation for $x_1(t)$ and the initial conditions are
\be  \label{15}
x^{\prime \prime}_1(t)+\omega^2 x_1(t)+D^2_1x_0(t)=0, 
\quad x_1(0)=0,  \quad x^{\prime}_1(0)=0,
\ee
where $D^2_1 x_0(t)$ is defined by (\ref{res1}):
\be \label{16}
D^2_1 x_0(t)=x^{\prime \prime}_0(0) \ln(t)+\gamma x^{\prime \prime}_0(t) +
\int^t_0 x^{(3)}_0(\tau) \ln(t-\tau) d \tau .
\ee
Substitution of Eq. (\ref{14}) into Eq. (\ref{16}) gives
\[ 
D_1x_0(t)=-\omega^2x(0) \ln(t)-
\gamma \omega^2 [x(0) \cos(\omega t)+(x^{\prime}(0)/\omega) \sin(\omega t)] +
\]
\be \label{17}
+\omega^3 x(0) \int^t_0 \sin(\omega \tau) \ln(t-\tau) d \tau -
\omega^2 x^{\prime}(0) \int^t_0 \cos(\omega \tau) \ln(t-\tau) d \tau.
\ee
The first and second integrals can be calculated, and we have  
the relations
\[
\int^t_0 \sin(\omega \tau) \ln(t-\tau) d \tau=
\]
\be \label{18}
=\omega^{-1}\left( \ln(t)+
\cos(\omega t)\left[\ln(\omega)+\gamma-\mathrm{Ci}(\omega t) \right] +
 \sin(\omega t)
\left[\frac{\gamma+1}{2}-\frac{\pi}{4}-\mathrm{Si}(\omega t) \right] \right) ,
\ee
and
\[
\int^t_0 \cos(\omega \tau) \ln(t-\tau) d \tau=
\]
\be \label{19}
=\omega^{-1} \left(
\cos(\omega t)\left[\frac{\gamma+1}{2}-\frac{\pi}{4}-\mathrm{Si}(\omega t) \right]+
\sin(\omega t) \left[-\ln(\omega)-\gamma+\mathrm{Ci}(\omega t) \right]  \right) .
\ee
Here $\mathrm{Ci}(t)$ and $\mathrm{Si}(t)$ are sine and cosine integral functions 
\cite{AS} respectively:
\[
\mathrm{Si}(t)=\int^t_0 \frac{ \sin(x)}{x} dx=\mathrm{si}(t)+\pi/2, \quad
\mathrm{Ci}(t)=-\int^{\infty}_t \frac{\cos(x)}{x}dx=-\mathrm{ci}(t) .
\]

As the result, Eq. (\ref{15}) transforms into
\[ x^{\prime \prime}_1(t)+\omega^2 x_1(t)+
\left[ \omega^2 x(0) \cos(\omega t)+\omega v_0 \sin(\omega t) \right]
\left(\ln(\omega)-\mathrm{Ci}(\omega t)\right)+ \]
\be \label{20}
+ \left[ \omega^2 x(0) \sin(\omega t)- \omega v_0 \cos(\omega t)\right]
\left( \frac{\gamma+1}{2}-\frac{\pi}{4}-\mathrm{Si}(\omega t) \right) =0. 
\ee
To simplify the calculations consider
$x(0)=1$, $x^{\prime}(0)=0$, and $\omega=1$. Then
$x_0(t)=\cos(t)$, and
\be \label{21}
x^{\prime \prime}_1(t)+x_1(t)-\cos(t) \mathrm{Ci}(t)- \sin(t) \mathrm{Si}(t)+
\left( \frac{\gamma+1}{2}-\frac{\pi}{4} \right) \sin(t)=0. 
\ee
The solution of this equation is
\[ x_1(t)= \cos(t) \int^t_0 \left[ 
\left( \frac{\gamma+1}{2}-\frac{\pi}{4} \right)[1-\cos^2(u)] 
- \sin(u)\cos(u) \mathrm{Ci}(u)- 
\mathrm{Si}(u)+\mathrm{Si}(u)\cos^2(u) \right] du+ \]
\be \label{22}
+ \sin(t) \int^t_0 \left[
-\left( \frac{\gamma+1}{2}-\frac{\pi}{4} \right) \sin(u)\cos(u) +
\cos^2(u) \mathrm{Ci}(u)+ \sin(u)\cos(u)\mathrm{Si}(u) \right] du ,
\ee
which after integration takes the form
\be \label{22b}
x_1(t)=-\frac{t}{2} \left[ \sin(t)\mathrm{Ci}(t)+\cos(t) \mathrm{Si}(t) \right] 
-\frac{1}{2}+\frac{1}{2} \cos(t).
\ee

Finally, Eq. (\ref{12}) provides the solution
\be \label{23}
x(t)=\cos(t)-\frac{\varepsilon}{2} \Bigl( 1-\cos(t)+
t \left[ \sin(t)\mathrm{Ci}(t)+\cos(t) \mathrm{Si}(t) \right] \Bigr)
+...
\ee 
The term, that is proportional to $\varepsilon$, can be considered
as a correction to the solution $x(t)=\cos(t)$ due to the 
fractional derivative of order $\alpha=2-\varepsilon$.

%%%%%%%%%

Let us compare this result with the exact solution for FLO.
The decomposition of (\ref{11a}) is \cite{GM2}:
%%We decompose the solution of Eq. (\ref{11}) into :
\be \label{ES1}
x(t)=x(0)\left[f_{\alpha,0}(t)+g_{\alpha,0}(t)\right]+
t x^{\prime}(0)\left[f_{\alpha,1}(t)+g_{\alpha,1}(t)\right],
\ee
where
\[ f_{\alpha,k}(t)=\frac{(-1)^k}{\pi} \int^{\infty}_0 e^{-rt}
\frac{r^{\alpha-1-k} \sin(\pi \alpha)}{
r^{2\alpha}+2r^{\alpha} \cos(\pi \alpha)+1} dr,
\]
\be \label{ES5}
g_{\alpha,k}(t)=\frac{2}{\alpha} e^{t \cos(\pi / \alpha)} \
\cos \left[ t \sin(\pi/\alpha)-{\pi k}/{\alpha } \right],
\quad (k=0,1) .
\ee
For the initial conditions $x(0)=1$, and $x^{\prime}(0)=0$:
\be \label{ES6}
x(t)=E_{\alpha} (-t^{\alpha})=
f_{\alpha,0}(t)+g_{\alpha,0}(t) .
\ee
Using $\alpha=2-\varepsilon$ and the condition $\varepsilon \ t \ll1$, 
we get 
\[
f_{\alpha,0}(t)=-\frac{\varepsilon}{2} 
\left( 1+ t \sin(t) \mathrm{Ci}(t)+t\cos(t)\mathrm{Si}(t) 
-t \frac{\pi}{2}\cos(t) \right) +..., \]
\be \label{fg}
g_{\alpha,0}(t)=\cos(t)+
\varepsilon \left(\frac{1}{2}-\frac{\pi}{4}t \right) \cos(t)+....
\ee
Substitution of (\ref{fg}) into (\ref{ES6}) gives Eq. (\ref{23}).

\subsection{Fractional oscillator in phase space}

The variety of definitions of fractional derivatives causes some 
inconvenience in their applications to different physical problems.
Particularly, this comment concerns the phase space 
definition for the fractional dynamics, which is not
uniquely defined in the general case.
As an example, consider
the fractional Caputo derivative of order
$\alpha=2-\varepsilon$ that can be defined as
\be \label{ss1}
D^{\alpha}=D^{2-\varepsilon}=J^{\varepsilon} D^{2},
\ee
where $J^{\varepsilon}$ is a fractional integration 
of order $\varepsilon$.
Therefore, we can use the relation  
\be \label{ss2}
D^{2-\varepsilon}=J^{\varepsilon} D^{2}=J^{\varepsilon} D^1 D^1=
(J^{\varepsilon}D^1) D^1=D^{1-\varepsilon} D^1 
\ee
to represent the fractional equation
\be \label{ss3}
D^{2-\varepsilon}x(t)=F(x(t))
\ee
as a system of two equations
\be \label{ss4}
D^1 x(t)=p(t), \quad D^{1-\varepsilon}p(t)=F(x(t)) ,
\ee
where first equation has derivative of integer order.

In general, we have an inequality 
%%%(\cite{SKM} Theorem 2.5)
\be \label{ss5}
D^{1-\varepsilon_1} D^{1-\varepsilon_2} \not= D^{2-(\varepsilon_1+\varepsilon_2)} ,
\ee
where $D^{1-\varepsilon}=J^{\varepsilon}D^1$,
i.e., the composition doesn't exist for the Caputo derivatives
\be
D^{1-\varepsilon_1} D^{1-\varepsilon_2} x(t)= 
D^{2-(\varepsilon_1+\varepsilon_2)}x(t)+
\frac{1}{\Gamma(\varepsilon_1+\varepsilon_2)} 
t^{\varepsilon_1+\varepsilon_2-1} x^{\prime}(0) ,
\ee
and the operations $D^1$ and $J^{\varepsilon}$ do not commute:
\be \label{RL2a}
D^1 J^{\varepsilon} x(t)=J^{\varepsilon} D^1 x(t)+
\frac{t^{\varepsilon-1}}{\Gamma(\varepsilon)} x(0) .
\ee

The symmetrized fractional equation
\be \label{ss6}
D^{1-\varepsilon/2} [D^{1-\varepsilon/2} x(t)]=F(x(t)),
\ee
is not the same as (\ref{ss3}),
since
\be
D^{1-\varepsilon/2} [D^{1-\varepsilon/2} x(t)]=
D^{2-\varepsilon} x(t) +
\frac{t^{\varepsilon-1}}{\Gamma(\varepsilon)} x^{\prime}(0) .
\ee
Eq. (\ref{ss6}) is equivalent to (\ref{ss3}) for $x^{\prime}(0)=0$.
Consider Eq. (\ref{ss6}), and represent it as a system of 
fractional equations
\be \label{ss9}
D^{1-\varepsilon/2} x(t)=p(t), \quad D^{1-\varepsilon/2}p(t)=F(x(t)) ,
\ee
where both equations are fractional equations.
Using Eq. (\ref{ss11}), we can rewrite (\ref{ss9}) as
\be \label{ss14}
x^{\prime}(t)+(\varepsilon/2) D^1_1x(t)+...=p(t), \quad 
p^{\prime}(t)+(\varepsilon/2) D^1_1p(t)+...=F(x(t)),
\ee
and consider their solution in the form 
\be \label{ss15}
x(t)=x_0(t)+\varepsilon x_1(t)+...., \quad
p(t)=p_0(t)+\varepsilon p_1(t)+...
\ee
As the result, we get a system of equations:
\[
x^{\prime}_0(t)=p_0(t), \quad p^{\prime}_0(t)=F(x_0(t)),
\]
\be \label{ss18}
x^{\prime}_1(t)+D^1_1 x_0(t)=p_1(t), \quad 
p^{\prime}_1(t)+D^1_1 p_0(t)=
\left(\frac{\partial F(x)}{\partial x} \right)_{x=x_0} x_1(t).
\ee
For the initial conditions $x_0(0)=1$, $p_0(0)=0$, and $F(x)=-x$,
the perturbation terms $D^1_1x_0$ and $D^1_1 p_0$ 
in (\ref{ss18}) are defined as
\[
D^1_1x_0(t)=-\gamma \sin(t)-\int^t_0 \cos(\omega \tau) \ln(t-\tau) d \tau,
\]
\be
D^1_1 p_0(t)=-\ln(t)-\gamma \cos(t) +
\int^t_0 \sin(\omega \tau) \ln(t-\tau) d \tau ,
\ee
and leads Eq. (\ref{ss18}) to the form
\[ 
x^{\prime}_1(t)+\cos(t)\mathrm{Si}(t)- \sin(t) \mathrm{Ci}(t)
-\left[\frac{\gamma+1}{2}-\frac{\pi}{4} \right] \cos(t)=p_1(t), 
\] 
\be
p^{\prime}_1(t)-\cos(t)\mathrm{Ci}(t)- \sin(t) \mathrm{Si}(t)+
\left[\frac{\gamma+1}{2}-\frac{\pi}{4} \right] \sin(t) =-x_1(t)
\ee
with solution
\[
x_1(t)=-t\sin(t) \mathrm{Ci}(t)-t\cos(t) \mathrm{Si}(t)+\cos(t)-1, \]
\be
p_1(t)=-\sin(t)\mathrm{Ci}(t)-\cos(t) \mathrm{Si}(t)
-t\cos(t)\mathrm{Ci}(t)+t\sin(t) \mathrm{Si}(t)-\sin(2t)-\sin(t).
\ee
This solution describes the phase space evolution of FLO 
%%%to attracted point $(x,p)=(0,0)$
for $\varepsilon t \ll 1$.

%%%%%%%%%%%%%%%%%%%%%%%%%%%%%%%%%%%%%%%%%
\subsection{Nonlinear fractional oscillator}

Let us consider a fractional nonlinear oscillator 
that is defined by the equation
\be  \label{24}
D^{2-\varepsilon} x(t)-x(t)+2x^3(t)=0,
\ee
where $0<\varepsilon \ll1$, and
$D^{2-\varepsilon}$ is a Caputo fractional derivative.
For the soliton-type solution
\be \label{27}
x_0(t)=\frac{1}{\cosh(t)}={\mathrm{sech}}(t) ,
\ee
the correction of the order $\varepsilon$ can be 
found similar to (\ref{23}):
\be  \label{28}
x^{\prime \prime}_1(t)+[6 {\mathrm{sech}}^2(t)-1]x_1(t)+D^2_1 {\mathrm{sech}}(t)=0, 
\ee
where 
\be \label{29}
D^2_1 {\mathrm{sech}}(t)=
-\ln(t)+\gamma \frac{\cosh^2(t)-2}{\cosh^3(t)} +
\int^t_0 \frac{\sinh^2(\tau)[\cosh^2(\tau)-6]}{\cosh^4(\tau)} \ln(t-\tau) d \tau .
\ee
The solution  $x_1(t)$ of equation (\ref{29}) 
can be derived numerically.
As the result we can get the corrections to the solution $x_0(t)$.

%%%%%%%%%%%%%%%%%%%%%%
\section{Asymptotic behavior for large $t$}

\subsection{Asymptotic representation of fractional derivative}

In the previous sections, we consider the intermediate time
asymptotic limited by the condition $\varepsilon t\ll1$.
In this section, we consider the opposite condition $t \rightarrow \infty$.

It is known \cite{C1,Podlubny}, that the Laplace transform 
of the Caputo fractional derivative  is
\be \label{33}
\int^{\infty}_0 e^{-st} \left[ D^{\alpha} x(t) \right]dt=
s^{\alpha} X(s) -\sum^{n-1}_{k=0} s^{\alpha-k-1} x^{(k)}(0) ,
\ee
where $n-1< \alpha \le n$, and $X(s)$ is a Laplace transform of $x(t)$:  
\be \label{33b}
X(s)=\int^{\infty}_0 e^{-st} x(t) dt .
\ee
Note that formula (\ref{33}) involves the initial conditions $x^{(k)}(0)$
with integer derivatives $x^{(k)}$. 
Therefore we can put the initial conditions in the usual way.
The functions $x(t)$ satisfy the condition
\be \label{33c}
\int^{\infty}_{0} e^{-st} |x(t)| < \infty.
\ee
For $1<\alpha\le 2$, Eq. (\ref{33}) has the form
\be \label{34}
\int^{\infty}_0 e^{-st} \left[ D^{\alpha} x(t) \right] dt=
s^{\alpha} X(s) -s^{\alpha-1} x(0) -s^{\alpha-2}x^{\prime}(0).
\ee
Inversion of (\ref{34}) gives
\be \label{35}
D^{\alpha} x(t)=
\frac{1}{2\pi i} \int_{Br} e^{st} \left[ s^{\alpha} X(s) 
-s^{\alpha-1} x(0) -s^{\alpha-2}x^{\prime}(0) \right] ds .
\ee
where $Br$ denotes the Bromwich contour, i.e., a line from 
$c-i\infty$ to $c+i\infty$, where $Re(s)=c$, and the contour  
is taken to the right of all singularities
in order to insure condition (\ref{33c}).
Closing the contour to the right will yield $x(t)=0$ for $t<0$.
For non-integer $\alpha$ the power function $s^{\alpha}$ is
uniquely defined as $s^{\alpha}=|s|^{\alpha} \mathrm{exp}[i \ \arg(s)]$,
with $-\pi < \arg(s) < \pi$, that is in the complex $s$-plane cut along 
the negative real semi-axis.

The asymptotic expansion ($\varepsilon t \gg 1$) 
can be formally obtained by the expanding $X(s)$ in powers of $s$, 
and then inverting term-by-term. 
The asymptotic for large $t$ corresponds to small $s$.
Consider the Taylor series 
\be \label{36}
X(s)=\sum^{\infty}_{k=0} \frac{1}{k!} X^{(k)}(0) s^k , 
\quad |s|<1 .
\ee
Substitution of (\ref{36}) into (\ref{35}) yields
\[
D^{\alpha} x(t)=
-x(0) \frac{1}{2\pi i} \int_{Br} e^{st} s^{\alpha-1}  ds 
-x^{\prime}(0) \frac{1}{2\pi i} \int_{Br} e^{st} s^{\alpha-2} ds + 
\]
\be \label{38}
+\sum^{\infty}_{k=0} \frac{1}{k!} X^{(k)}(0)
\frac{1}{2\pi i} \int_{Br} e^{st} s^{\alpha+k} ds .
\ee
Using the analytical continuation of  
\be \label{39}
\frac{1}{2\pi i} \int_{Br} s^{\beta} ds=
\frac{t^{-\beta-1}}{\Gamma(-\beta)}, 
\quad Re(\beta)<0 
\ee
to the half-plane $Re(\beta)\ge 0$,  
we arrive to the equation
\be
\label{40}
D^{\alpha} x(t)=-\frac{x(0) t^{-\alpha}}{\Gamma(1-\alpha)}
- \frac{x^{\prime}(0) t^{1-\alpha}}{\Gamma(2-\alpha)}
+\sum^{\infty}_{k=0}  
\frac{X^{(k)}(0) t^{-\alpha-k-1}}{k! \Gamma(-\alpha-k)}  ,
\quad (t \rightarrow \infty), \quad 1<\alpha<2
\ee
with the leading asymptotic term
\be \label{53}
D^{2-\varepsilon} x(t) \approx -x(0)\frac{t^{-2+\varepsilon}}{\Gamma(-1+\varepsilon)}
-x^{\prime}(0) \frac{t^{-1+\varepsilon}}{\Gamma(\varepsilon)}
\approx \varepsilon x(0) t^{-2+\varepsilon}-\varepsilon x^{\prime}(0) t^{-1+\varepsilon} ,
\quad (\varepsilon t \gg 1) . 
\ee
It is seen from (\ref{53}) that for arbitrary small $x^{\prime}(0)$
the first term can be neglected for sufficiently large $t$.

%%%%%%%%%%%
\subsection{Asymptotics for linear fractional oscillator}

Let us consider 
some applications of the results of the previous section.
For the linear fractional oscillator
\be \label{54}
D^{\alpha}x(t)+\omega^2 x(t)=0 ,
\ee
with $\alpha=2-\varepsilon$, we obtain from (\ref{53}), 
\be \label{56}
x(t) \approx \frac{x(0)}{\omega^2} \frac{t^{-2+\varepsilon}}{\Gamma(-1+\varepsilon)} +
\frac{x^{\prime}(0)}{\omega^2}\frac{t^{-1+\varepsilon}}{\Gamma(\varepsilon)}
\approx-\varepsilon \frac{x(0)}{\omega^2} t^{-2+\varepsilon}
+\varepsilon \frac{x^{\prime}(0)}{\omega^2} t^{-1+\varepsilon}, \quad
(\varepsilon t \gg 1).
\ee
This result can be also derived from
the asymptotic of the Mittag-Leffler function
using the exact solution of (\ref{54}), 
\be \label{57} 
x(t)=x(0)E_{\alpha} (-\omega^2 t^{\alpha})+
x^{\prime}(0) t E_{\alpha,2}(-\omega^2 t^{\alpha}) .
\ee

Let us use the following integral representation
\be \label{58}
E_{\alpha}(z)=\frac{1}{2\pi i}\int_{Ha} 
\frac{\xi^{\alpha-1} e^{\xi}}{\xi^{\alpha}-z} d\xi ,
\ee
where $Ha$ denotes the Hankel path, 
a loop which starts from $-\infty$ along the lower side
of the negative real axis, encircles the circular disc 
$|\xi|\le |z|^{1/\alpha}$ in the positive direction, and 
ends at $-\infty$ along the upper side of the negative real axis.
By the replacement $\xi^{\alpha} \rightarrow \xi$ Eq. (\ref{58})
transforms into \cite{Podlubny,GLL}:
\be \label{59}
E_{\alpha}(z)=\frac{1}{2 \pi i \alpha} 
\int_{\gamma(a,\delta)} \frac{e^{ \xi^{1/\alpha} } }{\xi -z} d\xi,
\quad (1<\alpha <2) ,
\ee
where $\pi \alpha/2 < \delta < min\{\pi, \pi \alpha\}$.
The contour $\gamma(a,\delta)$ consists of two rays
$S_{-\delta}=\{\arg(\xi)=-\delta, |\xi|\ge a\}$ and  
$S_{+\delta}=\{\arg(\xi)=+\delta, |\xi|\ge a\}$,
and a circular arc 
$C_{\delta}=\{|\xi|=1, -\delta \le arc(\xi) \le \delta \}$.
Let us denote the region on the left from $\gamma(a,\delta)$
as $G^{-}(a,\delta)$. Then \cite{GLL}:
\be \label{60}
E_{\alpha}(z)=-\sum^{\infty}_{n=1} 
\frac{z^{-n}}{\Gamma(1-\alpha n)}, \quad z \in G^{-}(a,\delta), 
\quad (|z| \rightarrow \infty),  
\ee 
and $\delta \le |\arg(z)|\le \pi$.
In our case, $z=-\omega^2 t^{\alpha}$, $\arg(z)=\pi$, and
\be \label{60b}
E_{\alpha}(-\omega^2 t^{\alpha})\approx
\frac{t^{-\alpha}}{\omega^2 \Gamma(1-\alpha)}
\approx - \varepsilon \frac{1}{\omega^2} t^{-2+\varepsilon} .
\ee
In a similar way, we get 
\be
E_{\alpha,2}(-\omega^2 t^{\alpha})\approx
\frac{t^{-\alpha}}{\omega^2 \Gamma(2-\alpha)}
\approx \varepsilon \frac{1}{\omega^2} t^{-2+\varepsilon} .
\ee

We arrive at the asymptotic result (\ref{56}) 
that exhibits an algebraic decay for $t \rightarrow \infty$.
This algebraic decay is the most important effect of 
the non-integer derivative in the considered fractional equations,
contrary to the exponential decay of the usual 
damped-oscillation and linear relaxation phenomena.

%%%%%%%%%%%%%%%%%%%%%%%%%%%%%%%%%%%%%%%%%%%%
\subsection{Nonlinear fractional oscillator}

The results of the previous section can be extended to the
nonlinear fractional oscillator.
Consider the equation
\be \label{46}
D^{2-\varepsilon}x(t)-x(t)+x^3(t)=0,
\ee
and try to find its solution in the form
\be
x(t)=x_0(t)+\varepsilon x_1 (t)+...
\ee
For $\varepsilon=0$, $x_0(t)$ satisfies the equation
\be \label{46b}
x^{\prime \prime}_0(t)-x_0(t)+x^3_0(t)=0 .
\ee 
Consider a particular solution
\be
x_0(t)=\sqrt{2} {\mathrm{sech}}(t) ,
\ee
and its Laplace transform \cite{BE}:
\be
X(s)=\int^{\infty}_{0} x_0(t) e^{-st} dt
=\frac{1}{\sqrt{2}} \Psi\left(\frac{s+3}{4}\right)-
\frac{1}{\sqrt{2}} \Psi\left(\frac{s+1}{4}\right),
\ee
where $\Psi(z)$ is the digamma function
\be
\Psi(z)=\frac{\Gamma^{\prime}(z)}{\Gamma(z)}=\frac{d}{dz} \ln[\Gamma(z)] .
\ee
For $z\ll1$
\be
\Psi(z)=-\frac{1}{z}-\gamma+\frac{\pi^2}{6}z+O(z^2), 
\ee
and therefore
\be
X(s)=\frac{\pi}{2}+\frac{1}{8}
\left[\Psi\left(1,\frac{3}{4}\right)-\Psi\left(1,\frac{1}{4}\right) 
\right] s+O(s^2), \quad (s\rightarrow 0) ,
\ee
where 
\be
\Psi(n,z) =\frac{d^n \Psi(z)}{dz^n},
\ee
is the $n$-th polygamma function, and  
\be
\frac{1}{8}\left[ \Psi\left(1,\frac{3}{4}\right)-
\Psi\left(1,\frac{1}{4}\right) \right]=-1.831931188...
\ee

Substitution of (\ref{53}) into Eq. (\ref{46}) gives
\be \label{46bb}
x^3(t)-x(t)- \frac{\sqrt{2}}{\Gamma(1-\alpha)}
t^{-\alpha} \approx 0 .
\ee
The leading asymptotic term is 
\be
x(t) \approx \frac{A(t)}{6 \Gamma(1-\alpha)}+\frac{2\Gamma(1-\alpha)}{A(t)},
\ee
where  
\be
A(t)=\left(108 \sqrt{2} \ t^{-\alpha}+ 
\left[-12 \Gamma^2(1-\alpha) +162 t^{-2\alpha}\right]^{1/2} \right)^{1/3} 
\approx  c_1+c_2 t^{-\alpha}
\quad (t \rightarrow \infty) ,
\ee
and
\be
c_1=2^{1/3}(-3\Gamma(1-\alpha))^{1/6} , \quad
c_2=2^{5/6} 18(-3\Gamma(1-\alpha))^{-1/3} .
\ee

%%%%%%%%%%%%%%%%%%%%%%%%%%%%%%%%%%%%%%%%%%%%%%%%
\section{Fractional Ginzburg-Landau (FGL) equation}

\subsection{Appearance of fractional derivatives in 
Ginzburg-Landau equation}

Let us recall the appearance of the nonlinear parabolic equation 
\cite{Leon,Light,Kad,ZS}, and 
the FGL suggested in Ref. \cite{Zaslavsky6} 
(see also \cite{Physica2005,Mil}).
Consider wave propagation in some media and 
present the wave vector $\bf k$ in the form
\begin{equation}
{\bf k} = {\bf k}_0 + {\bfkappa} = {\bf k}_0 + {\bfkappa}_{\parallel}
+ {\bfkappa}_{\perp}, 
\label{eq:36}
\end{equation}
where ${\bf k}_0$ is the unperturbed wave vector and subscripts
$(\parallel ,\perp )$ are taken respectively to the direction of ${\bf k}_0$. 
A symmetric dispersion law $\omega = \omega (k)$ for
$\kappa \ll k_0$ can be written as
\be \label{PE1}
\omega (k)= \omega (|{\bf k}|) \approx 
\omega (k_0)+ \ v_g \ (|{\bf k}| - k_0 )+
{1 \over 2} v^{\prime }_g \ (|{\bf k}| - k_0 )^2 ,
\ee
where
\be \label{PE2}
v_g = \left(\frac{\partial\omega}{\partial k} \right)_{k=k_0} , \quad
v^{\prime}_g=\left(\frac{\partial^2\omega}{\partial k^2}\right)_{k=k_0} ,
\ee
and
\be \label{PE3}
|{\bf k}|=|{\bf k}_0 + {\bfkappa}|=
\sqrt{({\bf k}_0+\kappa_{\parallel})^2+\kappa^2_{\perp}} \approx
k_0+\kappa_{\parallel}+\frac{1}{2k_0}\kappa^2_{\perp}.
\ee
Substitution of (\ref{PE3}) into (\ref{PE1}) gives
\be
\omega (k) \approx \omega_0 + v_g\kappa_{\parallel} +
{v_g\over 2k_0} \kappa_{\perp}^2 +
\frac{v^{\prime}_g}{2} \kappa_{\parallel}^2 ,
\label{eq:37}
\ee
where $\omega_0=\omega(k_0)$. 
Expression (\ref{eq:37}) in the dual space ("momentum representation") 
corresponds to the following equation in the coordinate space
\begin{equation}
i {\partial Z \over \partial t} =\omega_0 Z- 
i v_g {\partial Z \over \partial x}
 - {v_g \over 2k_0 } \Delta_{\perp} Z-
{v^{\prime}_g \over 2 } \Delta_{\parallel} Z
 \label{eq:38}
\end{equation}
with respect to the field $Z=Z(t,x,y,z)$, 
where $x$ is along ${\bf k}_0$, 
and we use the operator correspondence between the dual space 
and usual space-time:
\[
\omega (k)  \ \longleftrightarrow \
 i {\partial\over\partial t} , \quad
\kappa_{\parallel} \ \longleftrightarrow \ 
 -  i {\partial\over\partial x} , \] 
\be
{(\bfkappa}_{\perp})^2 \ \longleftrightarrow \ - \Delta_{\perp}=
- {\partial^2 \over\partial y^2 }-{\partial^2 \over\partial z^2 } , \quad
{(\bfkappa}_{\parallel})^2 \ \longleftrightarrow \ 
- \Delta_{\parallel}=- {\partial^2 \over\partial x^2 } .
\label{eq:39}
\ee
A generalization to the nonlinear case can be carried out similarly to
(\ref{eq:37}) through a nonlinear dispersion law dependence on the wave
amplitude:
\begin{equation} 
\omega = \omega (k,|Z|^2 ) \approx\omega (k,0) + b|Z|^2
 = \omega (|{\bf k}|) + b|Z|^2
 \label{eq:40}
\end{equation}
with some constant 
$b = \partial\omega (k, |Z|^2 )/\partial |Z|^2$ at $|Z| = 0$. 
In analogy to (\ref{eq:38}), we obtain from (\ref{eq:37}), 
and (\ref{eq:39}): 
\begin{equation}
i {\partial Z \over \partial t} =\omega(k_0) Z - 
i v_g {\partial Z \over \partial x}
 - {v_g \over 2k_0 } \Delta_{\perp} Z-
{v^{\prime}_g \over 2 } \Delta_{\parallel} Z+ b|Z|^2 Z .
 \label{eq:41}
\end{equation}
This equation is known as the nonlinear parabolic equation 
\cite{Leon,Light,Kad,ZS}. 
The change of variables from $(t,x,y,z)$ to $(t,x-v_gt,y,z)$ gives
\begin{equation}
-i {\partial Z \over \partial t}
 = {v_g \over 2k_0} \Delta_{\perp} Z +
{v^{\prime}_g \over 2} \Delta_{\parallel} Z 
- \omega(k_0) Z - b|Z|^2 Z 
 \label{eq:42}
\end{equation}
that is also known as the nonlinear Schr\"{o}dinger (NLS) equation.

Wave propagation in a media with fractal properties can be easily
generalized by rewriting the dispersion law (\ref{eq:37}), (\ref{eq:40})
in the following way \cite{Zaslavsky6}:
\begin{equation}
 \omega (k,|Z|^2 )= \omega (k_0 ,0) + v_g \kappa_{\parallel} +
g_1 ({\bfkappa}_{\perp}^2 )^{\alpha /2} +
g_2 ({\bfkappa}_{\parallel}^2 )^{\beta /2}+ b|Z|^2,
\quad (1<\alpha, \beta <2)
 \label{eq:43}
\end{equation}
with new constants $g_1$, $g_2$.

Using the connection between Riesz fractional derivative 
and its Fourier transform \cite{SKM} 
\begin{equation}
(-\Delta_{\perp} )^{\alpha /2} 
\longleftrightarrow
({\bfkappa}_{\perp}^2 )^{\alpha /2} ,
\quad
(-\Delta_{\parallel} )^{\beta /2} 
\longleftrightarrow  
({\bfkappa}_{\parallel}^2 )^{\beta /2} ,
 \label{eq:44}
\end{equation}
we obtain from (\ref{eq:43})
\begin{equation}
i {\partial Z \over \partial t} =- iv_g {\partial Z \over \partial x}
+ g_1 (-\Delta_{\perp} )^{\alpha /2}  Z 
+ g_2 (-\Delta_{\parallel} )^{\beta /2}  Z 
+ \omega_0 Z + b|Z|^2 Z,
  \label{eq:45}
\end{equation}
where $Z=Z(t,x,y,z)$. 
By changing the variables from $(t,x,y,z)$ to $(t,\xi,y,z)$, 
$\xi=x-v_gt$, and using 
\be
(-\Delta_{\parallel} )^{\beta /2}=
\frac{\partial^{\beta} }{\partial |x|^{\beta}}=
\frac{\partial^{\beta} }{\partial |\xi|^{\beta}}, 
\ee
we obtain  from (\ref{eq:45}) equation
\begin{equation}
i {\partial Z \over \partial t} =
 g_1 (-\Delta_{\perp} )^{\alpha /2}  Z 
+ g_2 (-\Delta_{\parallel} )^{\beta /2}  Z 
+ \omega_0 Z + b|Z|^2 Z,
\label{eq:46}
\end{equation}
that can be called the fractional nonlinear parabolic equation.
For $g_2=0$ we get the
nonstationary FGL equation (fractional NLS equation)
suggested in \cite{Zaslavsky6}.
Let us comment on the physical structure of (\ref{eq:46}).
The first term on the right-hand side is related to 
wave propagation in a media with fractal properties. 
The fractional derivative can also appear
as a result of ray chaos \cite{ZL,ref:16}
or due to a superdiffusive wave propagation 
(see also the discussion in \cite{Zaslavsky1,ZL}
and corresponding references therein). Other terms on the right-hand-side of
Eqs. (\ref{eq:45}), and (\ref{eq:46}) correspond to  wave interaction due to
the nonlinear properties of the media. Thus, Eq. (\ref{eq:46}) can describe
fractal processes of self-focusing and related issues.

We may consider one-dimensional simplifications of (\ref{eq:46}), i.e., 
\begin{equation}
i {\partial Z \over \partial t}=
g_2 \frac{\partial^{\beta} Z}{\partial |\xi|^{\beta}}
+ \omega_0 Z + b|Z|^2 Z,
\label{eq:47a}
\end{equation}
where $Z=Z(t,\xi)$, $\xi=x-v_gt$, or the equation
\begin{equation}
i {\partial Z \over \partial t} =
g_1 \frac{\partial^{\alpha} Z}{\partial |z|^{\alpha}}
+ \omega_0 Z + b|Z|^2 Z,
\label{eq:47b}
\end{equation}
where $Z=Z(t,z)$.
We can reduce (\ref{eq:47b}) to the case of a propagating wave 
\be
Z=Z(z-v_gt)\equiv Z(\eta) .
\ee
Then (\ref{eq:47b}) becomes
\be \label{eq:49}
g_1\frac{d^{\alpha}Z}{d |\eta|^{\alpha}}+c\frac{dZ}{d\eta}+\omega_0 Z+bZ^3=0,
\quad \eta=z-v_g t 
\ee
for real $Z(\eta)$, and $c=iv_g$.
This equation takes the form of a fractional generalization of
the Ginzburg-Landau equation (FGL), when $v_g=0$.
Eq. (\ref{eq:49}) differs from  the
fractional Burgers equation \cite{ref:14,ref:13}
in the structure of the nonlinear term. 
Nevertheless, an analysis similar to \cite{ref:14,ref:13}
may be performed to obtain some estimates on the solution 
\cite{Zaslavsky6}.

%%%%%%%%%%%%%%%%%%
\subsection{$\varepsilon$-expansion for the FGL equation}

Let us consider 
$\varepsilon$-expansion for a particular case of
equation (\ref{eq:47b}) when the time dependence
can be excluded by the replacement
\be \label{u1}
Z(t,z)=e^{i a t} \bar{Z}(z).
\ee
It gives the FGL equation
\begin{equation} \label{1D}
g\frac{d^{\alpha} \bar{Z}}{d |z|^{\alpha}} + 
\bar{\omega} \bar{Z} + b\bar{Z}^3=0, \quad
g=g_1, \quad \bar{\omega}=\omega_0+ac 
\end{equation}
for real field $\bar{Z}=\bar{Z}(z)$.
Again, let us search for a solution in the form
\be \label{GL2}
\bar{Z}(z)=\bar{Z}_0(z)+\varepsilon \bar{Z}_1(z)+... ,
\ee
and
\be \label{GL3}
\frac{d^{2-\varepsilon} \bar{Z}}{d |z|^{2-\varepsilon}}=
\frac{d^2 \bar{Z}}{d z^2} +\varepsilon D^2_1 \bar{Z}+...
\ee
similar to section 3.1.
In zero approximation $\bar{Z}(z)=\bar{Z}_0(z)$ which satisfies the equation
\be \label{GL4}
g\frac{d^2 \bar{Z}_0}{d z^2}+\bar{\omega} \bar{Z}_0 +
b \bar{Z}_0^3 =0.
\ee
As in section 3.3 consider a particular solution of (\ref{GL4}):
\be \label{GL12}
\bar{Z}_0(z)=
|2 \bar{\omega}/b|^{1/2} {\mathrm{sech}}
\left( |\bar{\omega}/g|^{1/2} \ z \right) .
\ee
In the first approximation
\be \label{GL5}
g\frac{d^2 \bar{Z}_1}{d z^2}+
(3b \bar{Z}^2_0+\bar{\omega}) \bar{Z}_1 +g D^2_{1}\bar{Z}_0=0,
\ee
%%%Note that this equation is linear with respect to $Z_1$.
where $\bar{Z}_0=\bar{Z}_0(z)$ is defined in (\ref{GL12}), and
\be \label{GL19}
D^2_{1}\bar{Z}_0(z)=\gamma \bar{Z}^{(2)}_0(z)+ 
\frac{1}{2} \int^{\infty}_{0} dy \ln(y)
[ \bar{Z}^{(3)}_0(z-y)-\bar{Z}^{(3)}_0(z+y)] .
\ee
The derivatives in (\ref{GL19}) are 
\[
\bar{Z}^{(2)}_0(z)=\left|\frac{2 \bar{\omega}}{b}\right|^{1/2}
\frac{A^2({\mathrm{cosh}}^2 (Az)-2)}{{\mathrm{cosh}}^3 (Az)}, \]
\be
\bar{Z}^{(3)}_0(z)=
\left|\frac{2 \bar{\omega}}{b}\right|^{1/2}
\frac{A^3{\mathrm{sinh}}(Az) 
(6-{\mathrm{cosh}}^2(Az)) }{{\mathrm{cosh}}^4(Az)},
\ee
where $A=|\bar{\omega}/g|^{1/2}$.
Eq. (\ref{GL5}) with respect to $\bar{Z}_1(z)$ has only
the integer derivatives and can be solved approximately or numerically.
The asymptotics $z\rightarrow \infty$ for (\ref{1D}) 
is discussed in Appendix 2.

\section{Conclusion}

Equations with fractional derivatives can be considered as 
a convenient tool to model different physical processes
with dissipation, persistent memory, and long 
range interaction. 
To work with such type equations, one needs different approximate
schemes such as perturbation theory, asymptotics expansions, etc.
In our paper we develop $\varepsilon$-expansion
for the cases when the order of fractional derivatives $\alpha$
can be presented as $\alpha=n-\varepsilon$, $n=1,2$, 
and $\varepsilon \ll 1$. This expansion is not uniform 
with respect to $t\gg 1$, and two cases $\varepsilon t\ll 1$,
$\varepsilon t \gg 1$, are very different.
We demonstrate  how these expansions work for some examples
that have exact a solution (fractional linear oscillator) 
and examples when the $\varepsilon$-expansion is not trivial
(fractional nonlinear oscillator, fractional Ginzburg-Landau
equation).
Fractional derivatives are not defined uniquely,
and for different types of derivatives the 
$\varepsilon$-expansion is different.

\section*{Acknowledgments}

Authors thank M.F. Shlesinger for valuable remarks.
This work was supported by the Office of Naval Research,
Grant No. N00014-02-1-0056, the U.S. Department
of Energy Grant No. DE-FG02-92ER54184, and the NSF
Grant No. DMS-0417800. 
VET thanks the Courant Institute of Mathematical Sciences
for support and kind hospitality.

%%%\newpage

\section*{Appendix 1: $\varepsilon$-expansion for
Riemann-Liouville fractional derivatives 
of order $\alpha=2-\varepsilon$ }

From the equation
\be \label{P2}
_x^CD^{\alpha}_b f(x)=\ _x{\cal D}^{\alpha}_b \left( f(x)-
\sum^{n-1}_{k=0} \frac{x^k}{k!} f^{(k)}(b)  \right) ,
\ee
where $n-1<\alpha<n$, follows
\be \label{RL2}
_0{\cal D}^{\alpha}_t f(t)=_0D^{\alpha}_t f(t)+
\sum^{n-1}_{k=0} \frac{t^{k-\alpha}}{\Gamma(k-\alpha+1)} f^{(k)}(0) ,
\ee
or for $n=2$, $\alpha=2-\varepsilon$
\be \label{RL4}
_0{\cal D}^{2-\varepsilon}_t f(t)=_0D^{2-\varepsilon}_t f(t)+
\frac{t^{-2+\varepsilon}}{\Gamma(-1+\varepsilon)} f(0)+
\frac{t^{-1+\varepsilon}}{\Gamma(\varepsilon)} f^{(1)}(0) .
\ee
For $\varepsilon \ t \ll1$, we use the expansions
\be \label{RL5}
\frac{t^{-2+\varepsilon}}{\Gamma(-1+\varepsilon)}=
t^{-2} \left(-\varepsilon+ \varepsilon^2[1-\gamma-\ln(t)]+... \right) ,
\ee
\be \label{RL6}
\frac{t^{-1+\varepsilon}}{\Gamma(\varepsilon)}=
t^{-1} \left(\varepsilon+ \varepsilon^2[\gamma+\ln(t)]+... \right) .
\ee
From Eq. (\ref{RL4}) and (\ref{7}), we obtain
\be \label{Ex1}
_0{\cal D}^{2-\varepsilon}_t f(t)
=f^{(2)}(t)  +\varepsilon {\cal D}^2_1 f +... ,
\ee 
where
\be \label{RL7}
{\cal D}^2_1 f(t)=-t^{-2} f(0)+ t^{-1} f^{(1)}(0)+
f^{(2)}(0) \ln(t)+ \gamma f^{(2)}(t)+ \int^t_0 f^{(3)}(\tau) \ln(t-\tau ) 
d \tau .
\ee 
Expressions (\ref{Ex1}), (\ref{RL7}) provides the $\varepsilon$-expansion
for Riemann-Liouville fractional derivative.

\section*{Appendix 2: Large $x$ asymptotics for FGL equation}

In section 4, we consider the asymptotic $t \rightarrow \infty$
for Caputo derivative.
Let us consider a modified equation (\ref{eq:47b}),
where we replace the Riesz derivative 
${\partial^{\alpha} Z}/{\partial |z|^{\alpha}}$ 
by the Caputo derivative $D^{\alpha} Z$:
\be \label{A1}
i\frac{\partial Z}{\partial t}=gD^{\alpha} Z+\omega_0Z+b |Z|^2 Z,
\ee
where $Z=Z(t,z)$, $g=g_1$, 
and the boundary condition is $Z^{(1)}_z(t,0)=0$. 
Again, exclude the term with first derivative, 
as in (\ref{u1}), by the replacement
\be
Z(t,z)=e^{i a t} \bar{Z}(z) .
\ee
It gives modified FGL equation
\be \label{ZZ1} 
gD^{\alpha} \bar{Z} + \bar{\omega} \bar{Z} + b|\bar{Z}|^2\bar{Z}=0 ,
\ee
where $\bar{\omega}=\omega_0+ac$, and $\bar{Z}^{(1)}(0)=0$.
Substitution of (\ref{53}) into (\ref{ZZ1}) yields
\be \label{z1} 
b|\bar{Z}|^2\bar{Z}+ \bar{\omega} \bar{Z}
-g\frac{z^{-\alpha}}{\Gamma(1-\alpha)} \bar{Z}(0) \approx 0.
\ee
Using representation
\be
\bar{Z}(z)=R(z) e^{i\phi(z)},
\ee
where $R(z)=|\bar{Z}(z)|$ is real and $\phi(z)=\arg (\bar{Z})$, Eq. (\ref{z1}) 
can be rewritten as
\be
bR^3(z)+\bar{\omega}R(z)-
g\frac{z^{-\alpha}}{\Gamma(1-\alpha)} R(0) e^{i[\phi(0)-\phi(z)]} \approx 0 .
\ee
If the constants $b$, $\bar{\omega}$, $g$ are real, then
\be \label{ZZ2}
bR^3(z)+\bar{\omega}R(z) \pm g\frac{z^{-\alpha}}{\Gamma(1-\alpha)} R(0) \approx 0,
\quad \phi(z)=\phi(0)+\pi n,
\ee
where $n$ is an integer.
The solution of (\ref{ZZ2}) is
\be
R_1(z)\approx \pm \varepsilon \frac{ gR(0) }{\bar{\omega} \Gamma(1-\alpha)} z^{-\alpha}
\approx \mp \varepsilon (g/ \bar{\omega}) R(0) z^{-2+\varepsilon} , \quad
(z \rightarrow \infty)
\ee
for $\bar{\omega} \not=0$, or
\be
R_2(z) \approx \pm
\frac{g^{1/3} z^{-\alpha/3}}{[b\Gamma(1-\alpha)]^{1/3}} R^{1/3}(0)  \approx
\mp \varepsilon^{1/3} (g/b)^{1/3} z^{(-2+\varepsilon)/3} R^{1/3}(0) 
\ee
for $\bar{\omega} =0$, $b\not=0$.
Finally, 
\be Z_1(t,z)
\approx \pm \frac{g z^{-\alpha} e^{i a t}}{\bar{\omega} \Gamma(1-\alpha)}  
Z(0,0) \approx 
\mp \varepsilon (g/\bar{\omega}) z^{-2+\varepsilon} e^{i a t} Z(0,0), 
\ee
and
\be Z_2(t,z) \approx 
\pm \frac{g^{1/3} z^{-\alpha/3} e^{i a t}}{
[b\Gamma(1-\alpha)]^{1/3}} R^{-2/3}(0) Z(0,0) \approx 
\mp  \varepsilon^{1/3} (g/b)^{1/3} z^{(-2+\varepsilon)/3} R^{-2/3}(0)
e^{i a t} Z(0,0). 
\ee

%%%\newpage
%%%%%%%%%%%%%%%%%%%%%%%%%%%%

\end{document}